 \documentclass[prd,preprintnumbers,amsmath,amssymb,floatfix]{revtex4}

\usepackage{graphicx}
\usepackage{dcolumn}
\usepackage{bm}
\usepackage{amssymb}
\usepackage{epsfig}
\usepackage{color}

\newcommand{\ba}{\begin{eqnarray}}
\newcommand{\ea}{\end{eqnarray}}
\newcommand{\be}{\begin{equation}}
\newcommand{\ee}{\end{equation}}
\newcommand{\bdisplay}{\begin{displaymath}}
\newcommand{\edisplay}{\end{displaymath}}

\newcommand{\eq}[1]{Eq.\,(\ref{#1})}

\newcommand{\chii}{\chi_{{}_{\rm I}}}

\newcommand{\sigtot}{\sigma_{\rm tot}}

\begin{document}

\title{ Forward hadronic scattering at 8 TeV: predictions for the LHC.}  
\author{Martin~M.~Block}
\affiliation{Department of Physics and Astronomy, Northwestern University, 
Evanston, IL 60208}
\author{Francis Halzen}
\affiliation{Department of Physics, University of Wisconsin, Madison, WI 53706}
\date{\today}

\begin{abstract}The Large Hadron Collider (LHC)  recently started operating at 8 TeV. In this note, we update our  earlier LHC   forward hadronic scattering predictions \cite{physicsreports,update7, blackdisk}, giving new  predictions, including errors, for the  $pp$ total and inelastic cross sections, the  $\rho$-value,  the  nuclear slope parameter $B$, $d\sigma_{\rm el}/dt$, and the large gap survival probability  at  8 TeV.
 \end{abstract}

\maketitle


\section{Introduction} \label{sec:introduction} 
After a long  successful run  at 7 TeV,  the LHC recently started extensive running at 8 TeV. We had made hadronic forward scattering predictions for the full design energy (14 Tev) for the LHC; for details see the review article by M. Block  \cite{physicsreports}.  More recently, we had calculated forward scattering parameters at 7 TeV \cite{update7,blackdisk}. The purpose of this note is to make predictions for 8  TeV, including errors  due to model uncertainties. Two  separate models were combined to make these predictions, the first being the analyticity-constrained analytic amplitude model of Block and Halzen \cite{newfroissart} that saturates the Froissart bound \cite{froissart} and the second being the ``Aspen Model'', a revised  version of the eikonal model of Block, Gregores, Halzen and Pancheri \cite{aspenmodel} that now incorporates analyticity constraints. Although self-contained, we  purposely keep explanations very brief; for more details, see Ref. \cite{physicsreports,update7,blackdisk}.

\section{Models used for predictions } \label{sec:predictions} 
\subsection{The analytic amplitude model, $\sigma_{\rm tot}$ and $\rho$}
We make the most accurate predictions of  the forward $pp$ scattering properties,
\ba
\sigma_{\rm tot}&\equiv&{4\pi\over p} {\rm Im}f(\theta_L=0)\label{sigtot1}\\
\rho&\equiv&{{\rm Re}f(\theta_L=0 )\over {\rm Im}f(\theta_L=0)},\label{rho}
\ea
using the analyticity-constrained analytic amplitude model of Block and Halzen \cite{newfroissart} that saturates the Froissart bound \cite{froissart}. By saturation of the Froissart bound, we mean  that the total cross section defined in \eq{sigtot1} rises as $\ln^2 s$, where $s$ is the square of the cms energy. In \eq{sigtot1} and \eq{rho}, $f(\theta_L)$ is the $pp$  laboratory scattering amplitude as a function of $\theta_L$, the laboratory scattering angle and $p$ is the laboratory momentum. In Fig. 1 of Ref. \cite{blackdisk}, where we showed that the proton asymptotically is a black disk of gluons whose radius goes as $R_0\ln(s)$, the upper dashed line (for $\sqrt(s)>100$ GeV) is  the total $pp$ ($\bar p p$) cross section as a function of the center-of-mass  energy, $\sqrt s$, given by
\ba
\sigma^0_{\rm tot}(\nu)&\equiv& 37.10\left(\frac{\nu}{m}\right)^{-0.5}+37.32-1.440\ln\left(\frac{\nu}{m}\right)+0.2817 \ln^2\left(\frac{\nu}{m}\right) \ {\rm mb},
\label{finaltotal}
\ea 
where $\nu$ is the laboratory energy and $m$ is the proton mass; thus $\nu/m\approx s/2m^2$. 
Our use of analyticity constraints---employing new Finite Energy Sum Rules (FESR) \cite {FESR}---allows us to use {\em very accurate  low energy cross section measurements to act as an anchor} that accurately fixes our high energy cross section predictions.  
At 8 TeV, we find that $\sigma_{\rm tot}=97.6\pm 1.1$ mb; see Ref. \cite{blackdisk}. The same set of parameters  predict $\sigma_{\rm tot}=107.3\pm 1.2$ mb at 14 TeV \cite{physicsreports}.  Further, using Ref. \cite{blackdisk},  we predict that $\rho_{pp}=0.134\pm 0.001$ at 8 TeV. 
\subsection{The``Aspen'' Model: an eikonal model for $pp$ scattering}
The ``Aspen'' model uses an unconventional  definition of the  eikonal $\chi(b,s)$ in impact parameter space $b$, so that 
\ba
\sigtot(s)&=&2\int \left[1-e^{-\chi_I(b,s)}\cos\left(\chi_R(b,s)\right)\right]\, d^2\vec b,\label{sigelofb20}\\
\rho(s)&=&\frac{\int e^{-\chi_I(b,s)}\sin(\chi_R(b,s))\,d^2\vec b}{\int\left[ 1-e^{-\chi_I(b,s)}\cos(\chi_R(b,s))\right]\,d^2\vec b}\quad, \label{rhoofb0}\\
B(s)&=&\frac{1}{2}\frac{\int | e^{-\chi_I(b,s)+i\chi_R(b,s)}-1|b^2\,d^2\vec b}{\int | e^{-\chi_I(b,s)+i\chi_R(b,s)}-1|\,d^2\vec b},\label{B}\\
\frac{d\sigma_{\rm el}}{dt}&=&\pi\left|\int J_0(qb)\left[ e^{-\chi_I(b,s)+i\chi_R(b,s)}-1\right]b\,db\right|^2,\label{dsdt}\\
\sigma_{\rm el}(s)&=&\int\left| e^{-\chi_I(b,s)+i\chi_R(b,s)}-1\right|^2\,d^2\vec b,\\
\sigma_{\rm inel}(s)&\equiv&\sigtot(s) -\sigma_{\rm el}(s)=\int \left( 1-e^{-2\chi_I(b,s)}\right)\,d^2\vec b,\label{siginel}
\ea 
where $\sigma_{\rm inel}(s)$ is the total inelastic cross section.
The even eikonal profile function $\chi^{ even}$, which is the only surviving term at the high energies considered here,  receives contributions 
from
quark-quark, quark-gluon and gluon-gluon interactions, and can be written in the factorized form
\begin{eqnarray}
\chi^{ even}(s,b) &=& \chi_{qq}(s,b)+\chi_{qg}(s,b)+\chi_{gg}(s,b)
\nonumber \\
&=& i\left [ \sigma_{qq}(s)W(b;\mu_{qq})
+ \sigma_{qg}(s)W(b;\sqrt{\mu_{qq}\mu_{gg}})
+ \sigma_{gg}(s)W(b;\mu_{gg})\right ]\, ,\label{chiintro}
\end{eqnarray}
where $\sigma_{ij}$ is the cross sections of the colliding partons, and
$W(b;\mu)$ is the overlap function in impact parameter space,
parameterized as the Fourier transform of a dipole form factor. The parameters $\mu_{qq}$ and $\mu_{gg}$ are masses which describe the ``area'' occupied by the quarks and gluons, respectively, in the colliding protons. In this model hadrons asymptotically evolve into black disks of
gluons \cite{blackdisk}. For details of the parameterization of the model, see Ref. \cite{physicsreports}.
\subsection{Total inelastic cross section, $\sigma_{\rm inel}$}

From \eq{siginel} and \eq{sigelofb20}, we calculate the {\em ratio} $r(s)=\sigma_{\rm inel}(s)/\sigma_{\rm tot}(s)$, because most errors due to  parameter uncertainties cancel in the ratio.  We then multiply $r(s)$ by the (more accurate) total cross section using \eq{sigtot1} (the analytic amplitude model) to obtain the inelastic cross section shown in Fig. 
1 of Ref . \cite{blackdisk}, as the lower (red) curve, given by
 \ba
\sigma^0_{\rm inel}(\nu)&\equiv& 62.59\left(\frac{\nu}{m}\right)^{-0.5}+24.09+0.1604 \ln\left(\frac{\nu}{m}\right)+ 0.1433 \ln^2\left(\frac{\nu}{m}\right) \ {\rm mb}.
\label{finalinelastic}
\ea 
At 8 TeV, we find $\sigma_{\rm inel}=70.4\pm 1.3$ mb.
\subsection{Elastic scattering, $B$ and $d\sigma_{\rm el}/dt$}
From \eq{B} we find that the nuclear slope parameter $B$, the logarithmic derivative of the elastic cross section  with respect to $t$ at $t=0$, is given by $B= 18.47\pm 0.12$ (GeV/c)$^{-2}$ at 8 TeV, where $t$ is the squared momentum transfer.   

At 8 TeV, using \eq{dsdt},  we plot the differential elastic scattering cross section $d\sigma_{\rm el}/dt$, in mb/(GeV/c)$^2$, against $|t|$, in (GeV/c)$^2$ as the solid (black) curve in Fig. \ref{fig:dsdt}.  Also shown is the approximation,  ${d\sigma \over dt}|_{t=0} e^{-B|t|}$,  valid for small $|t|$, which is the dashed (red) curve. The agreement is striking for small $t$. 

A few remarks are in order about Fig. \ref{fig:dsdt}.  It is clear from inspection that 
\ba
\sigma_{\rm el}&\equiv &\int_0^\infty {d\sigma_{\rm el}\over d|t|}d|t|\label{sigelexact}\\
&\approx&\int_0^\infty {d\sigma_{\rm el}\over d|t|}_{t=0} e^{-B |t|} d|t|\\
&=&{1\over B}{d\sigma_{\rm el}\over d|t|}_{t=0}, \label{sigelapprox}
\ea
a result that clearly does not depend on the details of where the dip in $|t|$ is located, nor on the value or shape of the high $|t|$ portion of ${d\sigma_{\rm el}/ d|t|}$, since these $|t|$ regions  contribute negligible amounts to the integral in \eq{sigelexact}. Our prediction at  7 TeV \cite{update7} for ${d\sigma_{\rm el}/ d|t|}_{t=0}$ was 476.0 mb/(GeV/c$^2$), whereas the Totem collaboration \cite{Totem1} measured $503.7\pm 26.7^{\rm syst}\pm 1.5^{\rm stat}$ mb/(GeV/c$^2$), in agreement within about one standard deviation. 

We here point out that the height of the first dip, i.e. the value of ${d\sigma_{\rm el}/ d|t|}$ at $|t|\approx 0.55$ GeV$^2$, is difficult to calculate because it results from the {\em interference} of the real and imaginary parts  of the scattering amplitude. The scattering amplitude at 7 TeV is dominated by its imaginary part; if there were {\em no} real part, there would be {\em no} interference and ${d\sigma_{\rm el}/ d|t|}$ in the dip would vanish reflecting a pure diffraction pattern given by the Bessel function $J_0(qb)$ in \eq{dsdt}.  As a result  a minimal change in the real part at the dip introduces a very large fractional change in the value of ${d\sigma_{\rm el}/ d|t|}$, as well as slightly changing its location in $|t|$. This is illustrated by the fact that none of  the models shown by the Totem collaboration {\cite{Totem2} accurately predict both the magnitude and the location of the first minimum; further, their anticipated value of ${d\sigma_{\rm el}/ d|t|}$ for the shoulder at larger $|t|$ varies widely, with none fitting the data. However, they are reasonably successful in reproducing  the slope $B$, as well as finding the approximate location of the first dip.
\begin{figure}[h,t,b] 
\begin{center}
\mbox{\epsfig{file=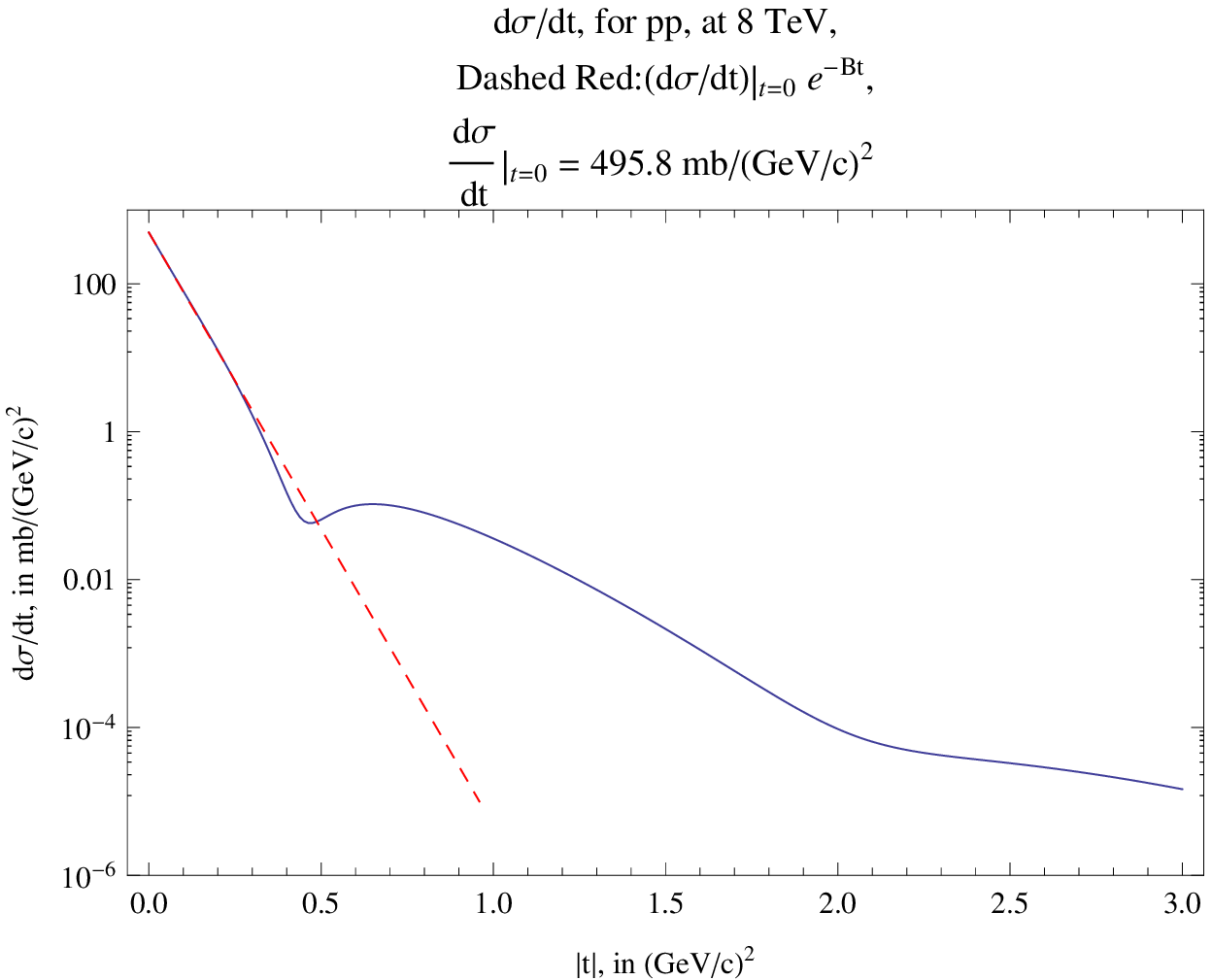
,width=3in%
,bbllx=0pt,bblly=0pt,bburx=371pt,bbury=229pt,clip=%
}}
\end{center}
\caption[]{
The 8 TeV  $pp$ differential elastic scattering cross section  , $d\sigma_{\rm el}/dt$, in mb/(GeV/c)$^2$, vs. $|t|$, in (GeV/c)$^2$ is the solid (black) curve. The dashed (red) curve is the small $|t|$ approximation, ${d\sigma \over dt}|_{t=0} e^{-B|t|}$.\label{fig:dsdt}
}
\end{figure}
\subsection{Rapidity gap survival probability}
As shown in Ref. \cite{physicsreports}, the survival probability $<|S|>^2$  of {\em any} large rapidity gap is given by
\be 
<|S|^2>=\int W(b\,;\mu_{\rm qq})\,e^{-2\chii(s,b)}d^2\,\vec{b},\label{eq:survival}
\ee 
which is the differential probability density in impact parameter space $b$ for {\em no} subsequent interaction (the exponential suppression factor)  multiplied by the quark probability distribution in $b$ space from \eq{chiintro}), which is then integrated over $b$.  It should be emphasized that \eq{eq:survival} is the probability of {\em survival} of a large rapidity gap and {\em not} the probability for the production and survival of large rapidity gaps, which is the quantity observed experimentally. The energy dependence of the survival probability $<|S|^2>$ is through the energy dependence of $\chii$, the imaginary portion of the eikonal given in \eq{chiintro}. A plot of $<|S|^2>$ as a function of $\sqrt s$, the cms energy in GeV, was given in Fig. 3 of Ref. \cite{update7}. At 8 TeV, we find the gap survival probability to be  $<|S|^2>=15.0\pm 0.05$ \%.
\section{Summary}We summarize our 8 TeV $pp$ forward scattering parameters for the LHC in Table \ref{table:LHC}, comparing them to our 14 TeV predictions taken from Ref. \cite{physicsreports}.

\begin{table}[h,t]                   
%
\def\arraystretch{1.5}            

\begin{center}
\caption[]{Values of forward scattering parameters for the LHC, at 8 and 14 TeV.
\label{table:LHC}
}
\vspace{.2in}
\begin{tabular}[b]{|c||c|c|c|c|c|}
\hline\hline
$\sqrt s$&$\sigma_{\rm tot}$&$\sigma_{\rm inel}$&$\rho$&$B$& $<|S|^2>$\\
(TeV)&mb&mb&&(GeV/c)$^{-2}$&\%\\
\hline
8&$97.6\pm 1.1$&$70.3\pm1.3$&$0.134\pm 0.001$&$18.47\pm 0.12$&$15.0\pm 0.05$\\
\hline
14&$107.3\pm 1.2$&$76.3\pm1.4$&$0.132\pm 0.001$&$19.39\pm0.13$&$12.6\pm0.06$\\
\hline\hline
\end{tabular}
\end{center}
\end{table}
\def\arraystretch{1}  
\begin{acknowledgments}
The work of F.H. is supported in part by the National Science Foundation under Grant No. OPP-0236449, by the DOE under grant DE-FG02-95ER40896 and in part by the University of Wisconsin Alumni Research Foundation.
One of us (M.M.B.) would like to thank  the Aspen Center for Physics, supported in part by NSF grant No. 1066293, for its hospitality during the time parts of this work were done.  
\end{acknowledgments}


\end{document}